\documentclass[twocolumn,showpacs,preprintnumbers,amsmath,amssymb]{revtex4}

\usepackage{graphicx}
\usepackage{dcolumn}
\usepackage{bm}

\begin{document}

\preprint{APS/123-QED}

\title{Quantum spin state stabilized by coupling with classical spins}

\author{Hironori Yamaguchi$^{1}$, Tsuyoshi Okubo$^{2}$, Akira Matsuo$^{3}$, Takashi Kawakami$^{4}$, Yoshiki Iwasaki$^{5}$, Taiki Takahashi$^{1}$, Yuko Hosokoshi$^{1}$, and Koichi Kindo$^{3}$}
\affiliation{
$^1$Department of Physics, Osaka Metropolitan University, Osaka 599-8531, Japan\\
$^2$Department of Physics, the University of Tokyo, Tokyo 113-0033, Japan\\
$^3$Institute for Solid State Physics, the University of Tokyo, Chiba 277-8581, Japan\\
$^4$Department of Chemistry, Osaka University, Toyonaka, Osaka 560-0043, Japan\\
$^5$Department of Physics, College of Humanities and Sciences, Nihon University, Tokyo 156-8550, Japan
}

\date{\today}

\begin{abstract}
We introduce a model compound featuring a spin-1/2 frustrated square lattice partially coupled by spin-5/2.  
A significant magnetization plateau exceeding 60 T could be observed, indicating a quantum state formed by $S$ = 1/2 spins in the square lattice.
The remaining $S$ = 5/2 spins exhibited paramagnetic behavior in the low-field regions.
The numerical analysis confirmed that the observed quantum state is a many-body entangled state based on the dominant AF interactions and is strongly stabilized by coupling with spin-5/2.
The stabilization of this quantum state can be attributed to a compensation effect similar to magnetic field-induced superconductivity, which serves as a strategy to control the stability of quantum spin states in magnetic fields.
\end{abstract}

\maketitle
Quantum spins in condensed-matter physics give rise to various entangled states, thus facilitating the studies on quantum computation achieved through single-qubit measurements on quantum spin states~\cite{MBQC,MBQC_rev}.
The main objective is to stabilize various quantum states in spin systems to serve as resources.
The gain of the magnetic energy through the formation of quantum states could be essential to quantize magnetic states in condensed-matter systems.
As a representative example, frustrated spin systems with competing exchange interactions induce quantum states to lower the ground state energy by suppressing the development of the magnetic moment. 
Extraordinary quantum states, such as the quantum spin liquid and the spin nematic, are proposed in low-dimensional frustrated systems~\cite{SL0,kita0,ne1}. 
If there is no frustration, slight perturbations, such as lattice distortion~\cite{dimer1, dimer2, dimer3} and bond randomness~\cite{uematsu}, in 2D systems can easily stabilize quantum states.
Even in one-dimensional (1D) systems with strong quantum fluctuations, highly entangled nonmagnetic quantum states are preferentially formed. 
The singlet state in the antiferromagnetic (AF) spin-1/2 chain with the lattice alternation, which corresponds to the dominant correlation in the present study, has lower energy than the ground state in the uniform AF chain.
In some 1D chain compounds, such singlet states can be formed even by distorting the lattice at a certain temperature, leading to a spin-Peierls transition~\cite{SP_exp1,SP_exp2,SP_exp3,SP_exp4,mySP}.

The flexibility of molecular orbitals in organic radical systems can generate advanced spin-lattice designs, leading to realization of a variety of quantum spin states. 
Our previous attempts in spin-lattice design using verdazyl-based quantum organic materials (V-QOM) have resulted in the formation of diverse quantum spin systems, including ferromagnetic-leg ladders~\cite{3Cl4FV,3Br4FV,3IV}, quantum pentagon~\cite{pentagon}, honeycomb with randomness~\cite{random}, and frustrated square lattices~\cite{square1,square2,square3,square4}, which were not achievabe with conventional organic and inorganic materials.
These spin models formed through V-QOM lead to unconventional quantum states that originate from highly entangled spins.
The next stage in spin-lattice design using V-QOM involves the synthesis of verdazyl-based salts by combining cationized verdazyl radicals with magnetic anions. 
This approach is aimed at achieving more diverse quantum spin states through the coupling between $\pi$ electrons in the radicals and 3$d$ electrons in the anion molecules ($\pi$-$d$ interaction).
This coupling results in magnetic field-induced quantum behavior~\cite{FeCl4,FeCl4_Br} and Lieb-Mattice ferrimagentic state~\cite{LM1,LM2,LM3}.
In the realm of $\pi$-$d$ systems, organic conductors with conducting $\pi$ electrons and localized $d$ spins were extensively investigated, particularly magnetic field-induced superconductivity~\cite{super_FeCl4, super_FeBr4}, wherein  the polarized $d$ spins stabilized the superconductivity by inducing an effective zero-field condition in the $\pi$ site~\cite{JP}. 
Similarly, in the present work, the quantum state formed in the radical salt is also expected to be stabilized through a similar mechanism through $\pi$-$d$ interactions.

In this letter, we present a model compound featuring a spin-1/2 spatially anisotropic frustrated square lattice partially coupled with spin-5/2. 
Single crystals of the verdazyl-based salt ($m$-MePy-V)$_2$MnCl$_4$ were successfully synthesized. 
Molecular orbital (MO) calculations revealed four types of exchange interactions, forming the spin-1/2 square lattice and partially coupling with spin-5/2. 
High-field measurements indicate a significant magnetic plateau exceeding 60 T, indicating the formation of a nonmagnetic quantum state by $S$ = 1/2 spins in the square lattice.
Numerical analysis demonstrates that this observed quantum state exhibits singlet-like correlation and is strongly stabilized by the coupling with spin-5/2.

To prepare $m$-MePy-V, we used a conventional procedure~\cite{gosei}, and ($m$-MePy-V)$_2$MnCl$_4$ was synthesized following a reported method for salts with similar chemical structures~\cite{square3, 3Dhoneycomb}.
The dark-brown ($m$-MePy-V)$_2$MnCl$_4$ crystals were obtained through recrystallization using methanol.
The crystal structure was determined from intensity data collected using a Rigaku AFC-8R Mercury CCD RA-Micro7 diffractometer. 
High-field magnetization measurements were carried out using a nondestructive pulse magnet, subjecting the samples to pulsed magnetic fields of up to approximately 60 T. 
The magnetic susceptibility was measured using a commercial SQUID magnetometer (MPMS-XL, Quantum Design). 
The experimental results were corrected considering the diamagnetic contributions calculated via Pascal's method. 
Specific heat measurements were performed using a commercial calorimeter (PPMS, Quantum Design). 
All experiments were conducted on small randomly oriented single crystals, considering the isotropic nature of organic radical systems.

The verdazyl radical $m$-MePy-V and MnCl$_4$ anion have spin-1/2 and 5/2, respectively~\cite{supple1}.
The crystallographic parameters at room-temperature are as follows: monoclinic, space group $P$2$_1$2$_1$2$_1$, $a$ = 10.6892(3) $\rm{\AA}$, $b$ = 15.2412(4) $\rm{\AA}$, $c$ = 24.4578(7) $\rm{\AA}$, $V$ = 3984.57(19) $\rm{\AA}^3$, $Z$ = 4, $R$ = 0.0389, and $R_{\rm{w}}$ = 0.1003. 
We performed MO calculations~\cite{MOcal} to evaluate the exchange interactions.
Subsequently, four types of dominant interactions were found between the $S$ = 1/2 spins on the radicals ~\cite{supple1}, as shown in Fig. 1(a).
They are identified as $J_{1}/k_{\rm{B}}$ = $72$ K, $J_{2}/k_{\rm{B}}$ = 45 K, $J_{3}/k_{\rm{B}}$ = 12 K, and $J_{4}/k_{\rm{B}}$ = $-8$ K, which are defined in the Heisenberg spin Hamiltonian given by $\mathcal {H} =
J_{n}{\sum^{}_{\langle i,j \rangle}}\textbf{{\textit
S}}_{i}{\cdot}\textbf{{\textit S}}_{j}$, where $\sum_{ \langle i,j \rangle}$ denotes the sum over the neighboring spin pairs.
Moreover, one dominant AF interaction, $J_{\rm{Mn}}$, is expected between spin-1/2 and 5/2~\cite{supple1}, as shown in Fig. 1(b).
Considering that the evaluation of absolute values of interactions between radical and MnCl$_4$ anion is difficult, we roughly evaluated as $J_{\rm{Mn}}$/$J_{1}$ $\simeq$ 0.2 using the following analysis of magnetic susceptibility.  
Consequently, $S$ = 1/2 spins form a spatially anisotropic square lattice through $J_{\rm{1}}$--$J_{\rm{4}}$ in the $ab$ plane, to which $S$ =5/2 spins are partially connected, as shown in Fig. 1(c). 
Each square unit has frustration caused by three AF interactions and one ferromagnetic interaction, which is divided into the following two patterns: $J_{\rm{1}}$-$J_{\rm{2}}$-$J_{\rm{3}}$-$J_{\rm{4}}$ and $J_{\rm{1}}$-$J_{\rm{3}}$-$J_{\rm{4}}$-$J_{\rm{2}}$.

\begin{figure*}[t]
\begin{center}
\includegraphics[width=38pc]{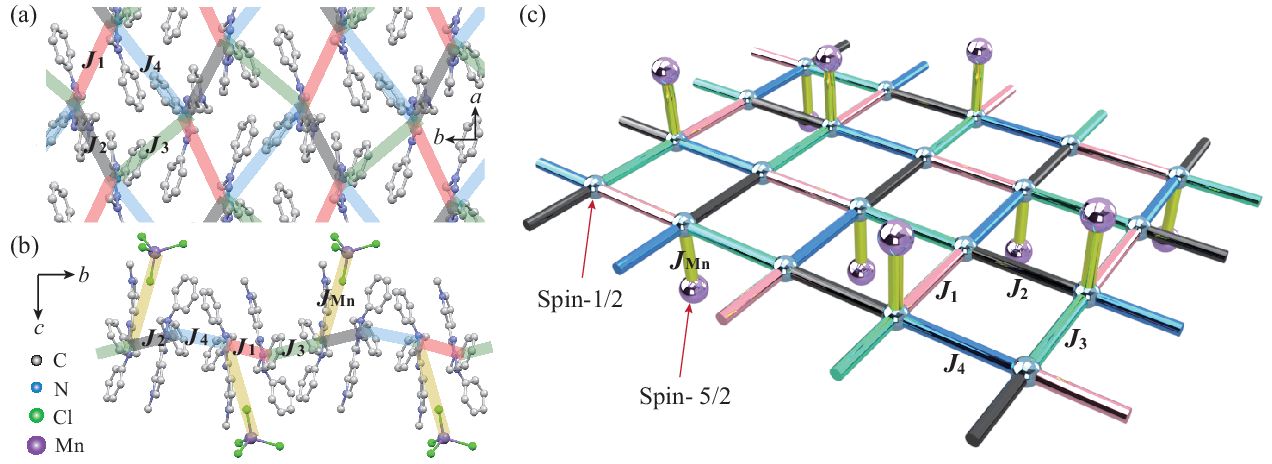}
\caption{(color online) (a) Crystal structure of ($m$-MePy-V)$_2$MnCl$_4$ in the (a) $ab$ and (b) $bc$ planes. Hydrogen atoms are omitted for clarity. The broken lines represent the bonds forming the spin lattice. 
(c) Two-dimensional spin model in the $ab$ plane. $J_{1}$, $J_{2}$, $J_{3}$, and $J_{4}$ form the frustrated square lattice of $S$ = 1/2 spins, and one of two $S$ = 1/2 sites is coupled by $S$ = 5/2 spins through $J_{\rm{Mn}}$.
}
\end{center}
\end{figure*}

Figure 2(a) shows the temperature dependence of the magnetic susceptibility ( $\chi$ = $M/H$) at 0.1 T, revealing a monotonic paramagnetic increase with decreasing temperatures. 
In the high-temperature region, $\chi$$T$ exhibits a rounded minimum at approximately 20 K, as shown in the inset of Fig. 2(a). 
This behavior is typical of mixed spin systems~\cite{LM1,mix1,mix2,mix3}, where interactions between spins of different sizes lead to excited states composed of smaller effective spins than those in the ground state, resulting in a rounded minimum in $\chi$$T$.  
In our system, the $\chi$$T$ behavior strongly depends on the value of $J_{\rm{Mn}}$ between spin-1/2 and -5/2.
Based on MO calculations that indicate $J_{\rm{1}}$ to be the most dominant interaction between $S$ =1/2 spins, we assumed a spin-(1/2,1/2,5/2) trimer coupled by $J_{\rm{1}}$ and $J_{\rm{Mn}}$ with the spin Hamiltonian given by $ \mathcal {H} = J_{\rm{1}}\textbf{{\textit S}}_{\rm{V1}}{\cdot}\textbf{{\textit S}}_{\rm{V2}}+J_{\rm{Mn}}\textbf{{\textit
S}}_{\rm{V2}}{\cdot}\textbf{{\textit S}}_{\rm{Mn}}$, where $\textbf{{\textit
S}}_{\rm{V}i}$ and $\textbf{{\textit S}}_{\rm{Mn}}$ are the spin-1/2 and spin-5/2 operators, respectively. 
The calculated results, as shown in the inset of Fig. 2(a), well reproduce the rounded minimum, allowing us to evaluate $J_{\rm{Mn}}$/$J_{\rm{1}}$ $\simeq$ 0.2.

Figure 2(b) displays the magnetization curve at 4.2 K measured in pulsed magnetic fields, with its inset showing the low-field region measured in static magnetic fields.
The observed hysteresis in pulsed magnetic fields originates from the slow relaxation characteristic of organic radical systems.
The magnetization gradually increases up to a value of 5 ${\mu}_{B}/\rm{f.u.}$, which persists up to at least 60 T.
Considering the isotropic $g$ values ($\sim$2.0) of the verdazyl radical and Mn$^{2+}$ ion, the saturation value is expected to be 7 ${\mu}_{B}/\rm{f.u.}$.
Therefore, the magnetic moment of 5 ${\mu}_{B}/\rm{f.u.}$ corresponds to the full polarization of spin-5/2 along the field direction, representing 5/7 of the full saturation value. 
The gradual increases in the low-field region can be effectively explained by the Brillouin function for spin-5/2 monomer, as shown in Fig. 2(b).
These behaviors suggest that the $S$ = 1/2 spins in the square lattice form a nonmagnetic quantum state.

The temperature dependence of the specific heat $C_{\rm{p}}$ is shown in Fig. 3(a). 
No peak indicative of a phase transition to a magnetic order is observed, but a significant change occurs below approximately 15 K when magnetic fields are applied.
As the low-temperature magnetization exhibits spin-5/2 paramagnetic behavior, the strong magnetic field dependence of $C_{\rm{p}}$ is considered to originate from the Zeeman splitting of spin-5/2, leading to Schottky contributions when magnetic fields are applied.
To evaluate the Schottky-type specific heat, we assumed Zeeman energy given by -$g{\mu}_{B}S_{z}$, where $S_z$ = $\pm$5/2, $\pm$3/2, $\pm$1/2.  
Furthermore, the magnetic specific heat $C_{\rm{m}}$ was determined by subtracting the lattice contribution $C_{\rm{l}}$ from the experimental results.
Below 15 K, $C_{\rm{l}}$ was approximated as $C_{\rm{l}}=a_{1}T^{3}+a_{2}T^{5}+a_{3}T^{7}$, a form that has been confirmed to be effective for verdazyl-based compounds~\cite{a235Cl3V, Zn_ferro}.
The constants $a_{1}-a_{3}$ were determined to reproduce the calculated Schottky-type specific heat.
As shown in Fig. 3(b), the magnetic specific heats at 3 T and 6 T are well explained by the calculated results assuming $C_{\rm{l}}$ with $a_{1}=0.022$, $a_{2}=$-$5.6\times10^{-5}$, and $a_{3}=5.1\times10^{-8}$.
The value of $a_{1}$ corresponds to a Debye temperature of 44 K, consistent with those used for other verdazyl-based compounds~\cite{3Cl4FV, 3Br4FV, square3, a235Cl3V}.
The evaluated $C_{\rm{m}}$ at zero-field is quite small, further supporting that the observed specific heats mainly originate from the Zeeman splitting of spin-5/2.
Figure 3(c) shows the magnetic entropy $S_{\rm{m}}$ obtained via the integration of $C_{\rm{m}}/T$, where the experimental data for 3 and 6 T have been shifted up by 4.7 and 1.2 J/ mol K, respectively.
The calculated results exhibit an asymptotic behavior toward the total magnetic entropy of spin-5/2 ($R$ln6 $\simeq$ 14.9).
Although the experimental temperature is not sufficient to observe the entire entropy shift, the temperature dependence of the experimental results is well reproduced by the entropy change from the higher temperature region.

\begin{figure}[t]
\begin{center}
\includegraphics[width=18pc]{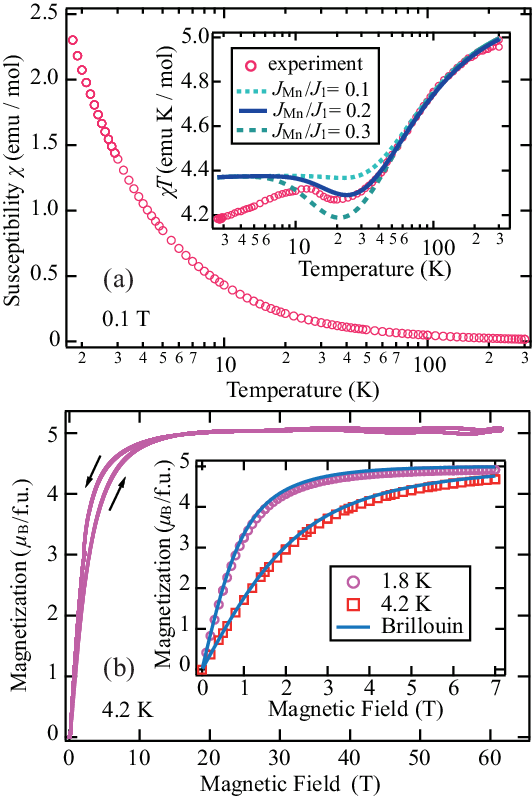}
\caption{(color online) (a) Temperature dependence of the magnetic susceptibility ($\chi$ = $M/H$) of  ($m$-MePy-V)$_2$MnCl$_4$ at 0.1 T. 
The inset shows the temperature dependence of $\chi T$.
The lines represent the calculated results for the spin-(1/2,1/2,5/2) trimer coupled by $J_{\rm{1}}$ and $J_{\rm{Mn}}$.
(b) Magnetization curves of ($m$-MePy-V)$_2$MnCl$_4$ at 4.2 K in pulsed magnetic fields.
The inset shows low-field region measured in static magnetic fields at 1.8 and 4.2 K.
The solid lines represent the Brillouin function for spin-5/2.}\label{f2}
\end{center}
\end{figure}

\begin{figure}[t]
\begin{center}
\includegraphics[width=21pc]{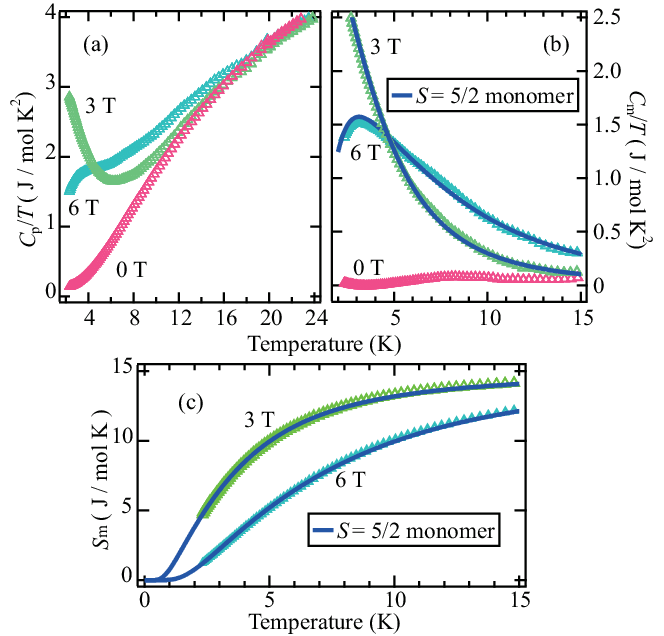}
\caption{(color online) (a) Temperature dependence of the specific heat $C_{\rm{p}}/T$ of ($m$-MePy-V)$_2$MnCl$_4$ at 0, 3  and 6 T. (b) Low-temperature region of $C_{\rm{m}}/T$.
The solid lines represent the Schottky-type specific heat originating from the Zeeman splitting of spin-5/2.
(c) Temperature dependence of the magnetic entropy $S_{\rm{m}}$.
The experimental values for 3 and 6 T have been shifted up by 4.7 and 1.2 J/ mol K, respectively, to overlap with the calculated results. 
}\label{f3}
\end{center}
\end{figure}

Here, we considered the ground state of the present system.
Our previous studies have demonstrated that MO calculations for verdazyl-based compounds reliably provide values of exchange interactions between radical spins, thus enabling the examination of their intrinsic behavior~\cite{3Cl4FV, 3Br4FV, 3IV, pentagon, square1, square2}.
Accordingly, considering the MO evaluations, we assumed the exchange interactions as $J_{2}/J_{1}$ = 0.63 and $J_{3}/J_{1}$ = 0.17 K.
Additionally, for the quantum Monte Carlo calculation~\cite{QMC}, we fixed $J_{4}$ = 0 to avoid the negative sign problem, leading to a non-frustrated case.
Regarding $J_{\rm{Mn}}$, we assumed $J_{\rm{Mn}}/J_{1}$ = 0.2, as evaluated from the magnetization analysis above.
We confirm that the 5/7 magnetization plateau appears up to approximately $H/J_{1}$ = 1.15, as shown in Fig. 4(a). 
Consequently, the observed plateau up to 60 T indicates that the actual value of $J_{1}$ is more than 70 K.  
Given that the magnetic properties below the plateau phase can be described by the spin-5/2 monomer, the plateau is considered to form an AF quantum state with fully polarized spin-5/2 along the field direction.
As the dominant AF interaction $J_{1}$ plays a crucial role and contributes more significantly than the other interactions, we anticipate the formation of a singlet-like quantum state in the 2D plane, as schematically illustrated in Fig. 4(a).

Finally, we examine the stability of the quantum state in association with frustration and coupling with spin-5/2.
Frustrated systems tend to lower their ground-state energies through lattice decouplings~\cite{dimen1,dimen2,dimen3,dimen4,dimen5}.
In the present lattice, the weakest ferromagnetic interaction $J_{4}$ is expected to be decoupled to minimize the energy loss due to frustration, thus stabilizing the singlet-like quantum state.
Moreover, ferromagnetic interactions are known to enhance the decoupling of the AF quantum state~\cite{square4, hagiwara1}, which aligns well with the $J_{4}$ decoupling scenario.
Accordingly, although it is difficult to quantitatively verify the precise effects of frustration, we can expect that the frustration acts to stabilize the quantum state in the present spin lattice. 
In terms of the effects of the coupling with spin-5/2, we calculated $J_{\rm{Mn}}/J_{1}$ dependence of the magnetization curves with fixed $J_{2}/J_{1}$ = 0.63 and $J_{3}/J_{1}$ = 0.17 K, as shown in Fig. 4(b).
We confirm that a case with no spin-5/2 coupling leads to a significant reduction in the plateau region.
These findings indicate that the quantum state in the square lattice is highly stabilized by coupling with spin-5/2.
This stabilization of the quantum state can be understood as a compensation effect, similar to the Jaccarino-Peterin mechanism for the magnetic field-induced superconductivity~\cite{super_FeCl4, super_FeBr4,JP}, where the external magnetic field compensates the internal magnetic field caused by the $\pi$-$d$ interaction.
In our case, the internal magnetic field induced by the fully-polarized spin-5/2 through $J_{\rm{Mn}}$ compensates the external magnetic field in the plateau phase, resulting in the expansion of the plateau region with the singlet-like quantum state as $J_{\rm{Mn}}$ increases.

\begin{figure}[t]
\begin{center}
\includegraphics[width=19pc]{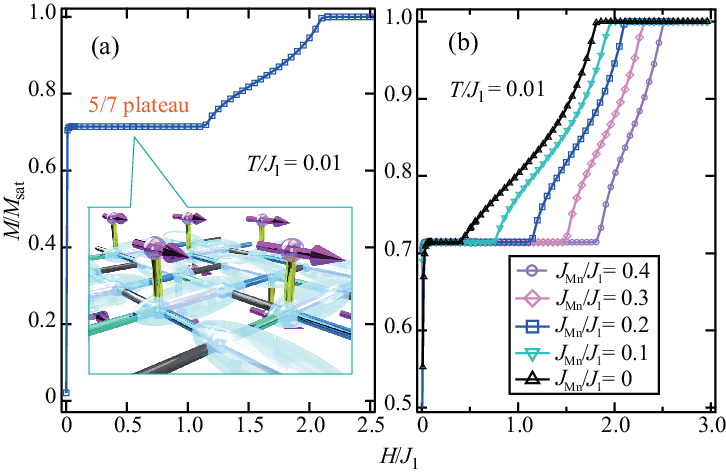}
\caption{(color online) (a) Calculated magnetization curve at $T/J_{1}$ = 0.01 for the expceted spin model assuming non-frustraed case ($J_{4}$ = 0) with $J_{2}/J_{1}$ = 0.63, $J_{3}/J_{1}$ = 0.17 K, and $J_{\rm{Mn}}/J_{1}$ = 0.2.
The illustrations describe the singlet-like quantum state protected by the fully-polarized spin-5/2 in the plateau phase.
(b) $J_{\rm{Mn}}/J_{1}$ dependence of the normalized magnetization curve at $T/J_{1}$ = 0.01.
}\label{f4}
\end{center}
\end{figure}

In summary, we succeeded in synthesizing single crystals of the verdazyl-based salt ($m$-MePy-V)$_2$MnCl$_4$. 
MO calculations indicated the formation of a spin-1/2 spatially anisotropic frustrated square lattice partially coupled with spin-5/2.
In the low-field region, the magnetizations and specific heats were well explained by spin-5/2 monomer that become fully polarized at approximately 10 T.
In the high-field measurement, we observed a large magnetization plateau exceeding 60 T, indicating a nonmagnetic quantum state formed by $S$ = 1/2 spins in the square lattice.
The numerical analysis demonstrated that the observed quantum state exhibits singlet-like correlations due to the dominant AF interaction.
Furthermore, we confirmed that the quantum state is strongly stabilized by coupling with spin-5/2, attributed to the compensation mechanism with the internal magnetic field induced by the fully-polarized spin-5/2.
Our study reveals a mechanism to stabilize quantum spin states in condensed materials and suggests a strategy to control the stability of quantum spin states in magnetic fields, which may inspire future technologies for quantum computation based on materials science.

\begin{acknowledgments}
We thank Y. Kono for valuable discussions.
This research was partly supported by the Murata Science Foundation and KAKENHI (Grants No. 23K13065 and No. 23H01127).
A part of this work was performed as the joint-research program of ISSP, the University of Tokyo and the Institute for Molecular Science.
\end{acknowledgments}



\begin{thebibliography}{99}
\bibitem{MOcal}
The MO calculations were performed using the UB3LYP method in the Gaussian 09 program package. 
The basis sets are 6-31G between radicals and 6-31G($d$,$p$) for $\pi$-$d$ couplings. 
For the estimation of intermolecular magnetic interaction, we applied the previously reported evaluation scheme~\cite{MO_method}.
\bibitem{QMC} 
The QMC code is based on the directed loop algorithm in the stochastic series expansion representation~\cite{QMC2}. 
The spin model unit contains eight radical sites with spin-1/2 and four Mn sites with spin-5/2.
The calculations was performed for $N$ = 768 under the periodic boundary condition, where $N$ denotes the system size.
It was confirmed that there is no significant size-dependent effect.
The number of sweeps is 50000 beginning after 10000 sweep thermalization.
All calculations were carried out using the ALPS application~\cite{ALPS,ALPS3}.
\end{thebibliography}
\end{document}